\documentclass[reprint,showpacs,preprintnumbers,pre,superscriptaddress]{revtex4-1}
\usepackage{graphicx}
\begin{document}

\title{Constitutive Relation for Nonlinear Response and Universality of Efficiency at Maximum Power for Tight-Coupling Heat Engines}
\author{Shiqi Sheng}
\affiliation{Department of Physics, Beijing Normal University, Beijing 100875, China}
\author{Z. C. Tu}\email[Corresponding author. Email: ]{tuzc@bnu.edu.cn}
\affiliation{Department of Physics, Beijing Normal University, Beijing 100875, China}


\begin{abstract}
We present a unified perspective on nonequilibrium heat engines by generalizing nonlinear irreversible thermodynamics. For tight-coupling heat engines, a generic constitutive relation for nonlinear response accurate up to the quadratic order is derived from the stalling condition and the symmetry argument. By applying this generic nonlinear constitutive relation to finite-time thermodynamics, we obtain the necessary and sufficient condition for the universality of efficiency at maximum power, which states that a tight-coupling heat engine takes the universal efficiency at maximum power up to the quadratic order if and only if either the engine symmetrically interacts with two heat reservoirs or the elementary thermal energy flowing through the engine matches the characteristic energy of the engine. Hence we solve the following paradox: On the one hand, the quadratic term in the universal efficiency at maximum power for tight-coupling heat engines turned out to be a consequence of symmetry [M. Esposito, K. Lindenberg, and C. Van den Broeck, Phys. Rev. Lett. \textbf{102}, 130602 (2009); S. Q. Sheng and Z. C. Tu, Phys. Rev. E \textbf{89}, 012129 (2014)]; On the other hand, typical heat engines such as the Curzon-Ahlborn endoreversible heat engine [F. L. Curzon and B. Ahlborn, {Am. J. Phys.} \textbf{43}, 22 (1975)] and the Feynman ratchet [Z. C. Tu, J. Phys. A \textbf{41}, 312003 (2008)] recover the universal efficiency at maximum power regardless of any symmetry.
\end{abstract}
\maketitle

\emph{Introduction.}--Energy-transduction devices such as heat engines~\cite{Curzon1975,Berry1977,devos85,Berry1985,Chen1989,ChenJC94,Bejan96,dcisbj2007,Esposito2010,GaveauPRL10,Espositopre2010,wangtu2012,Izumida2009PRE,Izumida2014PRL}, nano-motors~\cite{TuHu2005,LeighACIE07,FangHP07,HanggiRMP09}, and biological machines~\cite{Seifert30005,vdB210602,Seifert12rev,Imparato12,TuEPJE13} are crucial to our human activities. It is important to investigate their energetics in our times of resource shortages. Since they usually operate out of equilibrium, we need to develop some concepts of nonequilibrium thermodynamics to understand their operational mechanism.
Finite-time thermodynamics is a branch of nonequilibrium thermodynamics.
One of its most profound findings in recent years is the universality of efficiency at maximum power.
Up to the quadratic order of $\eta_{C}$ (the Carnot efficiency), the efficiencies at maximum power for the Curzon-Ahlborn endoreversible heat engine \cite{Curzon1975}, the stochastic heat engine~\cite{Schmiedl2008}, the Feynman ratchet~\cite{Tu2008}, and the quantum dot engine~\cite{Esposito2009EPL}, were found to coincide with a universal form
\begin{equation}\eta_{U}\equiv \eta_{C}/2+\eta_{C}^{2}/8+O(\eta_C^3),\label{eq-univeta}\end{equation}
where $O(\eta_C^3)$ represents the third and higher order terms of $\eta_C$.

The door towards this universality was opened by Van den Broeck~\cite{vdbrk2005} who proved that the linear term in (\ref{eq-univeta}) holds universally for tight-coupling heat engines working at maximum power. Next, considering a process of particle transport, Esposito \emph{et al.} found that the prefactor $1/8$ of the quadratic term in (\ref{eq-univeta}) is universal for tight-coupling heat engines in the presence of left-right symmetry~\cite{Esposito2009PRL}. This finding was confirmed by other nonlinear models of heat engines~\cite{Izumida2012,ApertetPRE13,TuShengJPA13,TuShengPRE14}.
Nevertheless, two typical heat engines recover universal efficiency (\ref{eq-univeta}) in the absence of symmetry. First, the efficiency at maximum power for the Curzon-Ahlborn heat engine~\cite{Curzon1975} is irrelevant to specific model-dependent parameters, and so regardless of any symmetry. Second, in the extremely asymmetric case, one of the present authors~\cite{Tu2008} optimized the power of the Feynman ratchet, and he found that the efficiency at maximum power still equates universal form (\ref{eq-univeta}). Additionally, Seifert pointed out that the Feynman ratchet still recovers the universality in other asymmetric cases~\cite{Seifert12rev}. Ironically, it is the Curzon-Ahlborn heat engine and the Feynman ratchet that arouse the issue of universality of efficiency at maximum power, on which researchers found that the universality of the quadratic term in (\ref{eq-univeta}) is attributed to the presence of symmetry, while both engines operating at maximum power take universal efficiency (\ref{eq-univeta}) in the absence of symmetry. This paradox has always puzzled researchers since the relationship between the universality and symmetry was discovered by Esposito and his coworkers~\cite{Esposito2009PRL}.

We aim at solving the above paradox from irreversible thermodynamics, a relatively mature framework in nonequilibrium thermodynamics. Its core quantity, entropy production rate, may be expressed as the sum of products of generalized thermodynamic fluxes and forces. The relation between the fluxes and forces is called constitutive relation. Although irreversible thermodynamics has been developed for many years, there still exists a controversy surrounding the definition of the generalized thermodynamic flux related to the heat flowing through a heat engine. One proposal is the rate of heat absorbed from the hot reservoir~\cite{vdbrk2005}; another choice is the mean rate of heat absorbed from the hot reservoir and that released into the cold reservoir~\cite{Jarzynski}. The present authors resolved this controversy by introducing the weighted thermal flux in recent work~\cite{TuShengPRE14}. However, the generic constitutive relation remains unknown for nonequilibrium heat engines in the regime of nonlinear response. The quadratic terms of thermodynamic forces have not been fully addressed in the previous work~\cite{TuShengPRE14} since they
disappear in the constitutive relation for the engines symmetrically interacting with two reservoirs. Similarly, the symmetric situation is also the focus of the pioneer work by Esposito and his coworkers~\cite{Esposito2009PRL}.
In contrast to the symmetric situation, the quadratic terms of thermodynamic forces should appear in the constitutive relation under asymmetric situations. We believe that a proper characterization of the constitutive relation up to the quadratic order is the key to solving the above paradox. In this Letter, we present a unified perspective on nonequilibrium heat engines, and then derive a generic nonlinear constitutive relation up to the quadratic order for tight-coupling heat engines from the stalling condition and the symmetry argument. Based on this generic relation, we obtain the necessary and sufficient condition for the universality of efficiency at maximum power, and hence solve the aforementioned paradox.

\begin{figure}[htp!]
\includegraphics[width=7cm]{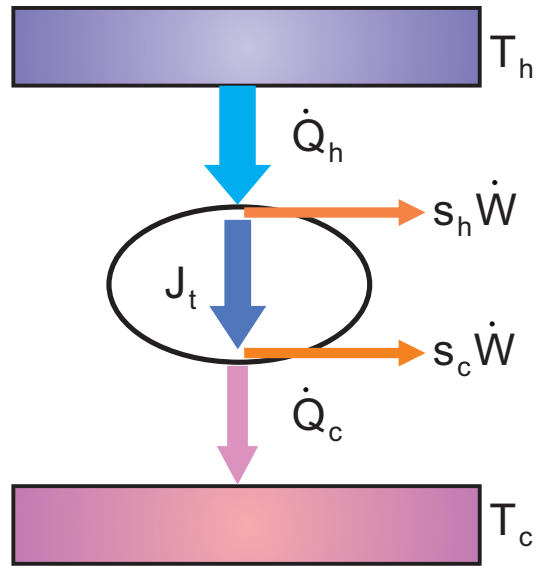} \caption{\label{fig1}(Color online) Refined generic model of a tight-coupling heat engine (reproduced according to Ref.~\cite{TuShengPRE14}).}
\end{figure}

\emph{Generic model.}--Above all, we briefly revisit a generic model for tight-coupling heat engines proposed in our previous work~\cite{TuShengPRE14}, which lays a solid theoretical foundation for the solution to the paradox. A heat engine may be simplified as the following schematic setup. The engine absorbs heat $\dot{Q}_{h}$ from the hot reservoir at temperature $T_{h}$, and releases heat $\dot{Q}_{c}$ into the cold reservoir at temperature $T_{c}$ per unit time. Simultaneously, it outputs a certain amount of power $\dot{W}$. By introducing two nonnegative weighted parameters $s_{h}$ and $s_{c}$ such that $s_{h}+s_{c}=1$, we define the weighted thermal flux
\begin{equation} J_{t} \equiv s_{h}\dot{Q}_{c}+ s_{c}\dot{Q}_{h},\label{Eq-model-Jt}\end{equation}
and the weighted reciprocal of temperature
\begin{equation} \beta \equiv {s_{h}}/{T_{h}}+{s_{c}}/{T_{c}}.\label{Eq-model-beta}\end{equation}
The values of weighted parameters $s_{h}$ and $s_{c}$ depend on specific models and they are related to the degree of symmetry of interactions between the heat engine and two reservoirs. In particular, $s_h=s_c =1/2$ indicates that the engine symmetrically interacts with two reservoirs. From definition (\ref{Eq-model-Jt}) and the energy conservation $\dot{Q}_{h}-\dot{Q}_{c}=\dot{W}$, we obtain $\dot{Q}_{h}=J_{t}+s_{h}\dot{W}$ and $\dot{Q}_{c}=J_{t}-s_{c}\dot{W}$, which lead to a refined generic model depicted in Fig.~\ref{fig1}. In this new physical picture, the engine absorbs heat $\dot{Q}_h$ per unit time from the hot reservoir, an amount of heat $s_h \dot{W}$ will be transformed into work output per unit time due to the interaction between the engine and the hot reservoir. A thermal flux $J_t$ flows through the heat engine, then an amount of heat $s_c \dot{W}$ will be transformed into work output per unit time due to the interaction between the engine and the cold reservoir. Finally, the engine releases heat $\dot{Q}_c$ per unit time into the cold reservoir. The contribution of interactions between the heat engine and the two reservoirs is explicitly included in this picture since the engine operates in a finite period or at a finite rate rather than in a quasistatic state. The reasonability of this picture and the significance of the weighted thermal flux were fully discussed in our previous work~\cite{TuShengPRE14}, which will not be repeated here.

The generalized thermal force conjugated to $J_{t}$ may be expressed as
\begin{equation} X_{t} \equiv 1/{T_{c}}-1/{T_{h}}.\label{Eq-model-Xt}\end{equation}
For a cyclic heat engine, the generalized mechanical flux $J_{m}$ and mechanical force $X_{m}$ may be defined as
\begin{equation} J_{m}\equiv 1/t_{0} \text{~and~} X_{m}\equiv -\beta W, \label{Eq-model-Jmc}\end{equation}
respectively, where $t_{0}$ is the period for completing the whole cycle. We emphasize that the sign of $t_{0}$ is of physical significance. $t_{0}$ takes a positive sign when the thermodynamic cycle corresponds to a genuine heat engine, while the negative sign represents the reverse cycle corresponding to a refrigerator.
For an autonomous heat engine operating in the steady state, the mechanical flux and mechanical force may be defined as
\begin{equation} J_{m}\equiv r \text{~and~} X_{m}\equiv -\beta w, \label{Eq-model-Jma}\end{equation}
respectively, where $r$ is the net rate and $w$ denotes the elementary work in each mechanical step.

With the consideration of definitions (\ref{Eq-model-Jt})--(\ref{Eq-model-Jma}), the entropy production rate $\sigma =
\dot{Q}_{c}/T_c-\dot{Q}_{h}/T_h$ of the whole system may be expressed as a canonical form $\sigma = J_{m}X_{m}+J_{t}X_{t}$. Let us focus on a tight-coupling heat engine, in which the heat-leakage vanishes so that the thermal flux is proportional to the mechanical flux,
\begin{equation}J_{t}/J_{m}=\xi,\label{Eq-model-tight-coupling}\end{equation}
where the ratio $\xi$ represents the elementary thermal energy flowing through the heat engine per thermodynamic cycle for a cyclic engine, or per spatial step for an autonomous engine. Then
the entropy production rate may be further expressed as $\sigma=J_{m}A$, where
\begin{equation}A\equiv X_{m}+\xi X_{t}\label{eq-affinity}\end{equation} is called affinity.
Particularly, $A=0$ represents a situation that the thermodynamic forces $X_{m}$ and $X_t$ balance each other. In this situation, the engine system is in a stalling state or quasistatic state with vanishing fluxes.

From (\ref{Eq-model-Jt}), (\ref{Eq-model-Jmc})-(\ref{Eq-model-tight-coupling}), we can derive the power output
\begin{equation} \dot{W}=-\beta^{-1}J_{m}X_{m}\label{Eq-model-W}\end{equation}
and the efficiency
\begin{equation} \eta=-X_{m}/(\beta \xi-s_{h}X_{m}).\label{Eq-model-efficiency}\end{equation}
Maximizing $\dot{W}$ with respect to $X_{m}$ for given $T_{c}$ and $T_{h}$, we obtain the optimization formula
\begin{equation} X_{m}(\partial J_{m}/\partial X_{m})+J_{m}=0.\label{Eq-model-Maximum}\end{equation}

\emph{Constitutive relation for nonlinear response.}--Now we generalize irreversible thermodynamics to the nonlinear regime by considering two essential arguments as follows.

First, we consider the stalling condition mentioned below (\ref{eq-affinity}) that $J_{m}$ should vanish when $A=0$. This condition requires $J_{m}$ to be formally expressed as
\begin{eqnarray} J_{m}=LA\left[1+v (A+u X_{t})\right]+O(A^{3},X_{t}^{3}),\label{Eq-solution-Jm1}\end{eqnarray}
where $L$, $v$ and $u$ are model-dependent coefficients. $O(A^{3},X_{t}^{3})$ represents the third and higher order terms of $A$ and $X_{t}$.

Second, we consider the contribution of symmetry by introducing an asymmetry parameter
$\lambda\equiv s_{h}-s_{c}$. The situation of $\lambda=0$ (i.e., $s_{h}=s_{c}=1/2$) corresponds to the case of symmetric interaction between the heat engine with two reservoirs. In this case, $J_{m}$ should be exactly reversed as all thermodynamic forces are reversed, which requires that all even-order terms in (\ref{Eq-solution-Jm1}) vanish, i.e., $v=0$ when $\lambda=0$. This requirement leads to $v=\alpha \lambda$ provided that $v$ is an analytical function, where $\alpha$ is a model-dependent parameter which could depend on $T_c$, $T_h$, $\lambda$ (or $s_h$), and so on. Substituting this equation into (\ref{Eq-solution-Jm1}), we transform $J_{m}$ into a generic form
\begin{equation}J_{m}=LA\left[1+\alpha \lambda (A+uX_{t})\right]+O(A^{3},X_{t}^{3}).\label{Eq-solution-Jm2}\end{equation}
For simplicity, the parameters $L$ and $\alpha$ in (\ref{Eq-solution-Jm2}) are respectively called the first and second master coefficients. This generic relation, as the first main result in this work, is uniquely determined from the stalling condition and the symmetry of system.

\emph{Necessary and sufficient condition.}--Now we address the efficiency at maximum power for a tight-coupling heat engine. By substituting (\ref{eq-affinity}) and (\ref{Eq-solution-Jm2}) into (\ref{Eq-model-Maximum}), we obtain the optimal mechanical force
$X_{m}^{\ast}=-\xi X_{t}/2+{\alpha \lambda\xi^{2}}X_{t}^{2}/{8}+O(X_{t}^{3})$.
Substituting it into (\ref{Eq-model-efficiency}) and considering (\ref{Eq-model-Xt}) and $\eta_C\equiv 1-T_c/T_h$, we finally achieve the efficiency at maximum power
\begin{equation} \eta^{\ast}=\frac{1}{2}\eta_{C}+\frac{1}{8}\eta_{C}^{2}+\frac{\lambda(1-\alpha\beta\xi)}{8}\eta_{C}^{2}+O(\eta_{C}^{3}),\label{Eq-solution-eta}\end{equation}
from which we obtain that the necessary and sufficient condition for the universal prefactor 1/8 of the quadratic term in (\ref{eq-univeta}) is $\lambda (1-\alpha\beta\xi)= O(\eta_{C})$. This condition may be further expressed as
\begin{equation}\lambda =0+O(\eta_{C})~\text{or}~\alpha\beta\xi=1+O(\eta_{C}).\label{Eq-solution-condition}\end{equation}

The physical meanings of (\ref{Eq-solution-condition}) are interpreted as follows. First, $\lambda =0+O(\eta_{C})$ is called symmetry condition, which represents that the heat engine interacts symmetrically with both heat reservoirs. Second, $\alpha\beta\xi=1+O(\eta_{C})$ is called energy-matching condition, which indicates that the elementary thermal energy ($\xi$) flowing through the heat engine matches the characteristic energy ($1/\beta$) of the heat engine since $1/\beta$ may be interpreted as the effective temperature~\cite{TuShengPRE14} of the heat engine and the Boltzmann constant has been set to unit. More precisely, the ratio of the characteristic energy of the heat engine to the elementary thermal energy flowing through the heat engine equals to $\alpha$, the second master coefficient of constitutive relation.

So far we get the second main result in the present work: Either the symmetry condition or the energy-matching condition results in universal efficiency (\ref{eq-univeta}) for tight-coupling heat engines working at maximum power. Indeed, it was proved that both the low-dissipation heat engine~\cite{Esposito2010,Schmiedl2008} and the minimally nonlinear irreversible heat engine~\cite{Izumida2012} take universal efficiency (\ref{eq-univeta}) when the symmetry condition is satisfied.
We conjecture that the reason why the Curzon-Ahlborn heat engine and the Feynman ratchet operating at maximum power recover universal efficiency (\ref{eq-univeta}) regardless of any symmetry is that the energy-matching condition is satisfied in both engines.

\emph{Curzon-Ahlborn heat engine.}--The Curzon-Ahlborn endoreversible heat engine \cite{Curzon1975} undergoes a cycle consisting of two isothermal processes and two adiabatic processes. In the isothermal expansion process, the working substance is in contact with a hot reservoir at temperature $T_{h}$. Its effective temperature is assumed to be $T_{he}$ ($T_{he}<T_{h}$). During time interval $t_{h}$, an amount of heat $Q_{h}$ is transferred from the hot reservoir to the working substance with the heat transfer law
\begin{equation}\label{eq-transQh}
Q_{h}=\kappa_{h}(T_{h}-T_{he})t_{h},\end{equation}
where $\kappa_{h}$ is the thermal conductivity in this process. The variation of entropy in this process is denoted by $\Delta S$. In the isothermal compression process, the working substance is in contact with a cold reservoir at temperature $T_{c}$. Its effective temperature is $T_{ce}$ ($T_{ce}>T_{c}$). During time interval $t_{c}$, an amount of heat $Q_{c}$ is transmitted from the working substance into the cold reservoir with the heat transfer law
\begin{equation}\label{eq-transQc}Q_{c}=\kappa_{c}(T_{ce}-T_{c})t_{c},\end{equation}
where $\kappa_{c}$ denotes the thermal conductivity in this process. The heat exchange and the entropy production are vanishing in the two adiabatic processes. The period ($t_{0}$) for completing the whole cycle is assumed to be proportional to $t_{c}+t_{h}$. In addition, the endoreversible assumption $Q_h/T_{he}=Q_c/T_{ce}$ is imposed on the engine.

According to equations (F2)--(F9) in Ref.~\cite{TuShengPRE14}, this engine may be mapped into the generic model. The main results are as follows:
\begin{equation}s_{h}=\frac{T_{h}\gamma_{c}}{T_{h}\gamma_{c}+T_{c}\gamma_{h}},~s_{c}=\frac{T_{c}\gamma_{h}}{T_{h}\gamma_{c}+T_{c}\gamma_{h}}\label{eq-CA4};\end{equation}
\begin{equation}\lambda\equiv s_h -s_c =\frac{T_{h}\gamma_{c}-T_{c}\gamma_{h}}{T_{h}\gamma_{c}+T_{c}\gamma_{h}},\end{equation}
\begin{equation}
J_{t}=T_{c}T_{h}\beta\Delta S J_{m}+O(J_{m}^3),\label{eq-CA6}\end{equation}
and
\begin{equation}J_{m}=\frac{\gamma_{c}\gamma_{h}}{(\gamma_{c}+\gamma_{h})\Delta S^{2}}A\left(1+\frac{1}{\Delta S} \lambda A\right)+O(A^{3},X_{t}^{3}),\label{Eq-CA-Jm}\end{equation}
with $\gamma_{h}\equiv\kappa_{h}t_{h}/t_{0}$, $\gamma_{c}\equiv \kappa_{c}t_{c}/t_{0}$, and $\lambda\equiv s_{h}-s_{c}=(T_{h}\gamma_{c}-T_{c}\gamma_{h})/(T_{h}\gamma_{c}+T_{c}\gamma_{h})$.
Obviously, (\ref{Eq-CA-Jm}) is a special form of generic expression (\ref{Eq-solution-Jm2}) with model-dependent parameters $L=\gamma_{c}\gamma_{h}/(\gamma_{c}+\gamma_{h})\Delta S^{2}$, $\alpha=1/\Delta S$ and $u=0$. In addition, equation (F6) in Ref.~\cite{TuShengPRE14} implies $\xi=T_{c}T_{h}\beta \Delta S$. Thus we obtain $\alpha\beta\xi=T_{c}T_{h}\beta^{2}=1+O(\eta_{C})$ with the consideration of (\ref{Eq-model-beta}), which conforms with the energy-matching condition in (\ref{Eq-solution-condition}).

\emph{Feynman ratchet.}--The Feynman ratchet~\cite{1Feynmanbook,1Buttiker,1Landauer} may be regarded as a Brownian particle walking in a periodic potential with a fixed step size $\theta$.
The Brownian particle is in contact with a hot reservoir at temperature $T_h$ in the left side of each energy barrier while it is in contact with a cold reservoir at temperature $T_c$ in the right side of each barrier. The particle moves across each barrier from left to right and outputs work against a load $z$. The height of energy barrier is $\epsilon$. The width of potential in the left or right side of the barrier is denoted by $\theta_{h}$ or $\theta_{c}=\theta-\theta_{h}$, respectively.
In the steady state and under the overdamping condition, according to the Arrhenius law \cite{1Feynmanbook}, the forward and backward jumping rates can be respectively expressed as
\begin{equation}\label{eq-Arrhenius}
R_{F} =r_{0}\mathrm{e}^{-(\epsilon +z\theta _{h})/T_h}, \mathrm{~~and~~} R_{B}=r_{0}\mathrm{e}^{-(\epsilon -z\theta _{c})/T_c},\end{equation} where $r_{0}$ represents the bare rate constant with dimension of time$^{-1}$.

The Feynman ratchet may be mapped into the refined generic model as shown in Ref.~\cite{TuShengPRE14}.
The main results are as follows:
\begin{equation}s_{h}=\theta_{h}/\theta,~s_{c}=\theta_{c}/\theta\label{eq-FR4};\end{equation}
\begin{equation}\lambda\equiv s_h -s_c =(\theta_{h}-\theta_{c})/\theta=(\theta_{h}-\theta_{c})/(\theta_{h}+\theta_{c}),\end{equation}
\begin{equation}
J_{t}=\epsilon J_{m},\label{eq-FR6}\end{equation}
and
\begin{equation}J_{m}=r_{0}\textrm{e}^{-\bar{\beta}\epsilon}A\left[1+\frac{\lambda}{2}(A-\epsilon X_{t})\right] +O(A^{3}, X_{t}^{3}),\label{Eq-Feynman-Jm}\end{equation}
where $\bar{\beta}=(1/T_{h}+1/T_{c})/2$.
Obviously, (\ref{Eq-Feynman-Jm}) is a specific form of generic expression (\ref{Eq-solution-Jm2}) with model-dependent parameters $L=r_{0}\textrm{e}^{-\bar{\beta}\epsilon}$, $\alpha=1/2$ and $u=-\epsilon=-\xi$.

In Ref.~\cite{Tu2008}, one of the present authors optimized the power of the Feynman ratchet with respect to both the external load $z$ and the internal barrier height $\epsilon$ under an extremely asymmetric situation ($\lambda=1$). He achieved the efficiency at maximum power $\eta^{\ast}=\eta_{C}/2+\eta_{C}^{2}/8+O(\eta_{C}^{3})$ and the corresponding optimal barrier height $\epsilon^{\ast}=T_{c}[1-\eta_{C}^{-1}\ln(1-\eta_{C})]=T_{c}[2+O(\eta_C)]$. Thus, we can easily verify
$\alpha\beta\xi=\beta\epsilon^{\ast}/2=1+O(\eta_{C})$
with the consideration of $\alpha=1/2$, $\xi=\epsilon^{\ast}$ and (\ref{Eq-model-beta}). In fact, for any case ($-1\le \lambda \le 1$), we can easily derive the corresponding optimal barrier height $\epsilon^{\ast}=T_c[(1-s_{h}\eta_{C})(1-\eta_{C})^{-1}-\eta_{C}^{-1}\ln(1-\eta_{C})]=T_{c}[2+O(\eta_C)]$ following the same optimization procedures as Ref.~\cite{Tu2008}. It is straightforward to verify $\alpha\beta\xi=\beta\epsilon^{\ast}/2=1+O(\eta_{C})$. Therefore, the Feynman ratchet always satisfy the energy-matching condition in (\ref{Eq-solution-condition}) when we optimize the power with respect to both the external load and the internal barrier height.

\emph{Conclusion.}--In summary, we dealt with nonequilibrium heat engines from a unified perspective and achieved the necessary and sufficient condition (\ref{Eq-solution-condition}) for the universality of efficiency at maximum power up to the quadratic order for tight-coupling heat engines. We found that both the Curzon-Ahlborn heat engine and the Feynman ratchet satisfy the energy-matching condition that guarantees universal efficiency (\ref{eq-univeta}) in the absence of symmetry. Hence we solved the paradox perfectly. More importantly, we phenomenologically wrote out generic nonlinear constitutive relation (\ref{Eq-solution-Jm2}) according to the stalling condition and the symmetry argument.
Such formula filled the knowledge gap in the literature and contributed substantially to nonequilibrium thermodynamics. This generic formula is well confirmed by typical models of heat engines such as the Curzon-Ahlborn heat engine, the Feynman ratchet mentioned above, and several examples illustrated in \cite{TuSheng15NJP}. Particularly, these models suggest that $\alpha$ in (\ref{Eq-solution-Jm2}) might be independent of the asymmetry parameter $\lambda$. We observe that these heat engines exhibit a kind of homotypy: The heat absorbed from the hot reservoir and that released into the cold reservoir by a cyclic heat engine abide by the same function type; The forward and backward flows for an autonomous heat engine also conform to the same function type. For these kind of heat engines, the second master coefficient $\alpha$ in (\ref{Eq-solution-Jm2}) is independent of $\lambda$ as shown in \cite{TuSheng15NJP}.

The present work may shed light on the future studies of nonequilibrium processes.
First, it is valuable if one can derive generic relation (\ref{Eq-solution-Jm2}) from statistical mechanics.
The application of fluctuation theorem~\cite{Seifert12rev,EvansFT2002,VWVdBE2014} in heat engines might be a starting point for this derivation. Second, low-dissipation heat engines~\cite{Esposito2010} and linear irreversible Carnot-like heat engines~\cite{wangtu2012} have the same bounds of efficiency at maximum power. It is possible to construct a connection between these two different types of heat engines within the present framework.

Finally, molecular motors~\cite{TuHu2005,LeighACIE07,FangHP07,HanggiRMP09,Seifert30005,vdB210602,Seifert12rev,Imparato12,TuEPJE13} in nano-world or biological realm look different from the heat engines in the above discussions. Most of them operate in a single heat reservoir and output work by utilizing the difference of chemical potentials rather than the temperature difference. By taking account of this distinction, we expect that the present unified perspective on nonequilibrium heat engines may be transplanted to understanding the optimization mechanism in energetics of molecular motors.

\emph{Acknowledgement.}--The authors are grateful to financial support from the
National Natural Science Foundation of China (Grant No. 11322543). They also thank Haiping Fang, Lamberto Rondoni, and Massimiliano Esposito, Qiangfei Xia for their instructive discussions and suggestions.

\end{document}